 \documentclass[final,5p,times,twocolumn,numbers]{elsarticle}

\usepackage{amssymb,amsmath}
\usepackage{graphicx}
\usepackage{tikz}
\usepackage{physics}
\usepackage{MnSymbol}
\usepackage{feynmp}
\usepackage{float}
\usepackage{tikz-feynman}

\usepackage[numbers]{natbib}
\usepackage{lipsum}
\usepackage{amssymb,amsmath}
\usepackage{graphicx}% Include figure files
\usepackage{tikz}
\usepackage{physics}
\usepackage{MnSymbol}
\usepackage{feynmp}
\usepackage{float}
\usepackage{graphicx}
\usepackage{tikz-feynman}
\tikzfeynmanset{
  graviton/.style={double, dashed},
}

\begin{document}

\begin{frontmatter}


%% \author{Victor H. Alencar\corref{cor1}\fnref{label2}}
%% \ead{victor.alencar@ufrj.br}

\title{Black Hole Evaporation as a Topological Tunneling}


\author[first]{Victor H. Alencar}
\ead{victoralencar@pos.if.ufrj.br}
\affiliation[first]{organization={Universidade Federal do Rio de Janeiro}, \newline %Department and Organization
            addressline={Instituto de Física --}, \newline
            city={Rio de Janeiro}, 
            postcode={}, \\  \newline
            state={RJ,}, \newline
            country={Brazil}}

\begin{abstract}
\par We present the quantization of the electromagnetic field near the event horizon of a Schwarzschild black hole using Euclidean path integrals. Our result for the vacuum energy describes a black hole surrounded by a finite volume of photons at $T_{H} = \frac{1}{8\pi G M}$, the black hole quantum atmosphere. The total entropy includes contributions from this atmosphere, and the Bekenstein entropy, which arises from the Gibbons--Hawking--York boundary term, which encodes topological information. We show that the contribution of the quantum atmosphere to the black hole specific heat is positive, indicating that the system may become thermodynamically stable.
\par By analyzing homology groups, we show that the black hole evaporation is a tunneling between topologically distinct spacetimes: Schwarzschild ($\chi = 2)$ transitions to the flat spacetime ($\chi = 1$) via Hawking radiation, where $\chi$ is the Euler characteristic, a topological invariant. This process resembles instanton-driven tunneling in Yang-Mills theories, where topologically non-trivial solutions dominate the vacuum amplitude. In our case, the Gibbons--Hawking--York term dominates the transition amplitude, which induces the evaporation process. These results corroborate the Parikh-Wilczek picture of Hawking radiation and the interpretation of Euclidean black holes as gravitational instantons.
\end{abstract}



\begin{keyword}
%% keywords here, in the form: keyword \sep keyword, up to a maximum of 6 keywords
Black Hole Thermodynamics  \sep Euclidean Path Integrals \sep Hawking Radiation \sep Quantum Atmospheres of Black Holes



\end{keyword}

\end{frontmatter}

%\tableofcontents



\section{Introduction}
\label{introduction}

\par Black holes are central in the pursuit of quantum gravity due to their extreme curvature, which enhances the interplay between quantum and gravitational physics. The main breakthrough in this area is the prediction of the emission of thermal radiation due to quantum fluctuations near the event horizon 
  \cite{Hawking:1974rv, Hawking:1975vcx}, the Hawking radiation. This mechanism is not exclusive to black holes; it is also present in cosmology 
\cite{GibbonsHawking1977, Hawking:1982dh, Polarski1989, Ciambelli:2020qny} and analog models of gravity  \cite{Unruh:1980cg, Barros:2020zkt, Mertens:2022jij, Steinhauer:2015saa}. The complete evaporation of black holes led to the \textit{information loss problem}, an ongoing debate \cite{Zhang:2025pzm, Pegas:2024abf, Chen:2014jwq, Almheiri2021, Zhang:2025xes} about the fate of information inside black holes.
\par Recently, a formula relating the Hawking temperature and the Euler characteristics, $\chi$ -- a topological invariant -- was discovered by Robson, Villanca and Bianchi (RVB) \cite{Robson:2018con}. The RVB formula applies to gravitational and synthetic event horizons. This discovery further confirms the topological origins of black hole thermodynamics \cite{GibbonsHawking1977, Eguchi:1978xp, Gibbons:1979xm, Liberati:1995jj, Liberati:1996kt, Liberati:1997sp, Ciambelli:2020qny, Hughes:2025ved, Volovik:2025wax}. Here we provide a new proof of the RVB formula by using the homology groups \cite{Hatcher,Schwarz:1994ee,Nakahara:2003} of black hole spacetimes. Our derivation extends the $2D$ formula to $D$-dimensional spacetimes. We show that algebraic methods \cite{Hatcher} are more suitable to study the topological effects of $D-$dimensional black holes than analytical techniques, such as the Chern-Gauss-Bonnet theorem. Because increasing the dimension of the spacetime leads to more complex integrals, in contrast, the algebraic calculations remain the same in any dimension due to the Künneth formula \cite{Hatcher, Schwarz:1994ee}. As an application of our generalization, we provide a new computation of the Hawking temperature of $D$-dimensional Schwarzschild-Tangherlini black holes \cite{Emparan:2008eg}. The generalized formula provides a nonlocal (global) method to evaluate the Hawking temperature of a spacetime.
\par Using Euclidean path integrals, we show that the vacuum amplitude near the event horizon is dominated by the Gibbons--Hawking--York boundary term, which yields the Bekenstein entropy of the black hole. The contribution of electromagnetic fluctuations near the horizon is expressed in terms of determinants of the Laplace--Beltrami operator, $g^{\mu\nu}\nabla_{\mu}\nabla_{\nu}$. The resulting partition function describes an evaporating black hole surrounded by an atmosphere of photons at the Hawking temperature $T_{H}=\frac{1}{8\pi G M}$. The existence of black hole quantum atmospheres has previously been argued using the canonical formalism \cite{Giddings:2015uzr, Hod:2016hdd, Dey:2017yez, Dey:2019ugf, Ong:2020hti, Zhang:2025xes}. Here, for the first time, we predict the formation of such atmospheres within a Euclidean path integral framework. Our expression for the vacuum energy indicates that matter fields near the event horizon behave similarly to ordinary thermal fields, in agreement with recent arguments by Biggs and Maldacena \cite{Biggs:2024dgp}. In addition, we compute the contribution of the quantum atmosphere to the total entropy of the black hole—which may be interpreted as entanglement entropy \cite{Witten:2024upt}—as well as its contribution to the specific heat. We find a positive correction, indicating that the presence of the quantum atmosphere may lead to a stabilization of the black hole.

\par In order to analyse the topological features of black hole evaporation, we computed the variation of the spacetime homology due to the evaporation. We show that this process is related to a variation in the $\chi$ of the spacetime: a stationary 4D black hole, $\mathbb{S}^{2}\times \mathbb{R}^{2}$ \cite{Hawking:1971vc,hawking1973large,galloway1993topology}, transitions to the flat spacetime, $\mathbb{R}^{4}$. As $H^{c}_{2}(\mathbb{S}^{2})\neq H^{c}_{2}(\mathbb{R}^{2})$, Schwarzschild and Minkowski spacetimes have differente Euler characteristics, $\chi(\mathcal{M}_{BH})=2$ and $\chi(\mathcal{M}_{\text{Flat}})=1$. The dominance of a boundary term in the vacuum amplitude and the change in a topological invariant are quite similar to tunnellings driven by instantons in Yang-Mills theories \cite{tHooft:1976snw, Callan:1976je, Shifman:1994ee, Shifman:2012zz}. Given these results, we can have a quantum interpretation of Euclidean black holes: it a class of solutions that describes tunnelings between topologically inequivalent spacetimes.
\par In addition, we show that our path integral calculation reproduces the results of the WKB derivation of Hawking radiation originally discovered by M. Parikh and F. Wilczek \cite{Parikh:1999mf} and subsequently extended by others \cite{Angheben:2005rm, Erbin:2017zwo}, further corroborating the interpretation of black hole evaporation as a tunneling process. We also propose that the Euler characteristic $\chi$ can be viewed as a quantum number of the black hole state, following the analogy between black holes and atoms initiated by J. Bekenstein \cite{Bekenstein:1974jk, Mukhanov:1986me}.

\section{Path Integral Quantization of the Electromagnetic Field in Curved Spacetimes}
\par The dynamics of the Einstein-Maxwell theory is given by the action,
\begin{equation}
S[A_{\mu};g_{\nu\lambda}] = \frac{1}{16\pi G}\int_{\mathcal{M}} d^{4}x\sqrt{-g}\  R - \frac{1}{4}\int_{\mathcal{M}} d^{4}x \sqrt{-g} \ F_{\mu\nu}F^{\mu\nu}  \nonumber 
\end{equation}
\begin{equation}
-  \frac{1}{8\pi G}\int_{\partial\mathcal{M}} d^{3}x\sqrt{-h} \ K,
\end{equation}
where the first two terms give the field equations. The last term is the Gibbons-Hawking-York term \cite{York1972, GibbonsHawking1977b}, a boundary term; hence, it does not affect the field equations. However, it is necessary to keep the action stationary and to reproduce the correct ADM energy \cite{Hartmann:2015,Witten:2024upt}.
\par The Einstein-Maxwell theory is not renormalizable \cite{Deser:1974cz,Deser:1974zzd} because general relativity is ill-defined at very short lengths \cite{Goroff:1985sz}. Given the lack of a quantum theory of gravity, a successful approach is semiclassical gravity; the quantization of fields in curved spacetimes \cite{Gibbons:1978dw,Wald:1995yp}.
\begin{figure}[H]
	\centering 
	\includegraphics[width=0.3\textwidth, angle=0]{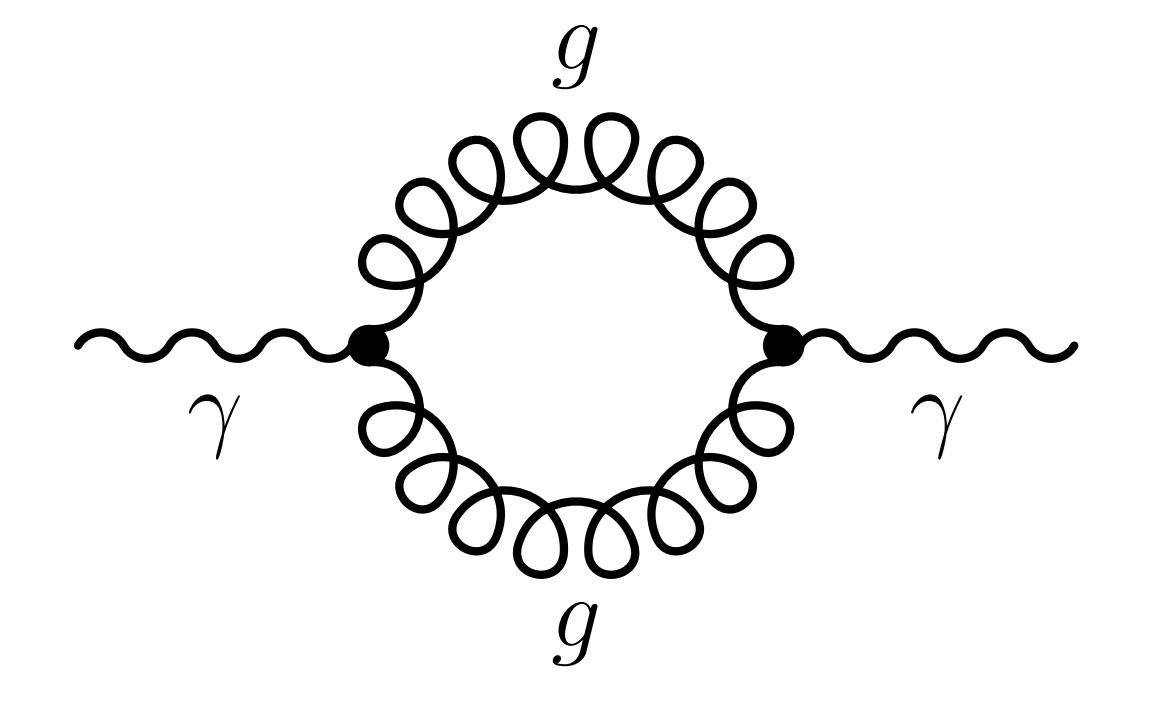}	
	\caption{This is one of the diagrams of the gravitational contribution to the electromagnetic vacuum polarization. As Maxwell-Einstein theory is not renormalizable, thus, it is not possible to obtain finite results to such processses. For this reason, here we will use quantum field theory in curved backgrounds, an effective approach to quantum gravity.} 
	\label{fig_mom0}%
\end{figure}

\par The quantum theory of fields in the absence of gravity heavily relies on the representation theory of the Poincaré group \cite{WeinbergI}, the isometry group of the Minkowski spacetime. In the presence of gravity, this becomes an issue as, in general, spacetimes do not possess isometry groups \cite{carroll2003spacetime,Hawking:1971vc}. Consequently, the vacuum in curved spacetimes is trickier than in flat ones, leading to issues such as a consistent definition of particles in curved spacetimes. To circumvent these problems, we can define the vacuum in the asymptotic regions of the spacetime, where we can apply the usual framework of quantum field theory.
\par Here we will use the Euclidean path integrals \cite{Gibbons:1978dw}, within this framework, we focused on the computation of thermodynamical observables, such as specific heat, which are independent of the concept of a single particle. Solving the path integral, we have the partition function that describes the thermodynamics of a black hole taking into account the electromagnetic vacuum.
\par The metric of the Schwarzschild black hole in spherical coordinates is
\begin{equation}
    ds^2 = g_{\mu\nu}dx^{\mu}dx^{\nu}= -\left(1 - \frac{r_{s}}{r} \right) dt^2 + \left(1 - \frac{r_{s}}{r} \right)^{-1} dr^2 + r^2 d\Omega^2,
\end{equation}
where $r_{s}=2GM$, the Schwarzschild radius. This line element has two singularities: $r=0$, a genuine singularity, and $ r=r_{s}$, which appears as a singular region due to the system of coordinates; in fact, this region is the event horizon \cite{carroll2003spacetime,hawking1973large}. The topology of the Schwarzschild black hole is $\mathbb{R}^{2}\times \mathbb{S}^{2}$ \cite{Gibbons:1979xm,galloway1993topology}, the same topology of the Reissner-Nordström, Kerr, and Kerr-Newman solutions \cite{Hawking:1971vc,hawking1973large,Jacobson:1994hs}. In general, the topology of asymptotically flat stationary black holes is $\mathbb{R}^{2}\times \mathbb{S}^{2}$.
\par The Wick rotation, $t \to it_{\textbf{E}}$, the analytic continuation to the Euclidean
 metric can be used to apply the dimensional regularization \cite{Bollini:1972ui}, a
 covariant and gauge invariant renormalization procedure \cite{costello2011renormalization}. From the mathematical perspective, Euclidean field theory is better defined because there is a Lebesgue-like measure theory for Euclidean path integrals \cite{simon2005functional}, allowing, for example, a formal proof of the renormalizability of Yang-Mills theories \cite{costello2011renormalization}. Currently, a rebirth of the Euclidean path integral approach to quantum fields in curved spacetimes in the framework itself \cite{Kontsevich:2021dmb, Witten:2018lgb, Witten:2021nzp,Visser:2021ucg} and applications to black holes and cosmology\cite{Hertog:2023vot, Held:2024qcl}.

\par In the Schwarzschild spacetime, the Wick rotation leads to the compactification of the Euclidean time coordinate, $t_{E} \sim t_{E} + 4\pi r_{s}$; this result can be found by computing the transition between spherical and Kruskal-Szekeres coordinates, as we show in \ref{Eu}. In quantum field theory, the periodicity of the Euclidean time coordinate indicates a thermal behaviour, in particular, a KMS state \cite{Svaiter:2008zz} with temperature
\begin{equation}
    T_{H} = \frac{1}{4\pi r_{s}} = \frac{1}{8\pi GM}.
\end{equation}
\par The periodic "Euclidean time" coordinate comes from the restrictions of the analytic continuation due to the topology of the Schwarzschild spacetime. Locally, the compactification is necessary to avoid a conical singularity \cite{Hartmann:2015,Witten:2024upt}. Globally, the conical singularity appears due to the event horizon, which is related to $\chi$ 
\cite{hawking1973large,Hawking:1971vc,Robson:2018con}.
\par The rich physics of the vacuum amplitude is crucial in a plethora of quantum systems, from Casimir effect to string theory. Here we will investigate quantum fluctuations near black holes by evaluating the vacuum amplitude using stationary phase and spectral methods \cite{gustafson2020mathematical, Percacci:2017fkn, Fujikawa, Hawking:1976ja}. 
\par The dynamics of the photons in the vicinity of a Schwarzschild black hole is given by Euclidean path integral,
\begin{equation}
    \mathcal{Z} = e^{-S_{GHY}[g_{\mu\nu}]}\int \mathcal{D}A_{\mu} \ e^{-\int \sqrt{-g}\big(\frac{1}{4}F_{\mu\nu}F^{\mu\nu}\big)},
\end{equation}
where the Hilbert-Einstein action does not contributes as $g^{\mu\nu}R_{\mu\nu}=0$, only the Gibbons-Hawking-York boundary term will contribute to the path integral,
\begin{equation}
    S_{GHY}[g^{Sch}] = -\frac{1}{8\pi G}\int_{\partial M} d^{3}y \sqrt{h} \ K,
\end{equation}
where $K$ is the trace of the extrinsic curvature, $K_{jk}=D_{j}n_{k}$, the covariant derivative of $n_{k}$, the unit vector normal to $\partial\mathcal{M}$. But the Schwarzschild spacetime, $\mathbb{R}^{2}\times\mathbb{S}^{2}$, is not compact the Gibbons-Hawking-York action diverges in $r$ \cite{Hartmann:2015,Witten:2024upt}. This infrared divergence is regularized by imposing a maximum finite value for the radial coordinate, $R_{M}$. Such regularization is implemented in the quantization procedure by introducing a renormalized Gibbons-Hawking-York term,
\begin{equation}
    S^{R}_{GHY}[g_{\mu\nu}] = \frac{1}{8\pi G} \int_{\partial \mathcal{M}} d^{3}y\sqrt{h} \ K +F\Big(\frac{1}{R_{M}}\Big) \nonumber
\end{equation}
\begin{equation}
    -\frac{1}{8\pi G} \int_{\partial \mathcal{M}} d^{3}y\sqrt{h} \ K_{0},
\end{equation}
where $F\Big(\frac{1}{R_{M}}\Big)$ is an analytic function that we will define below, and $K_{0}$ is the extrinsic curvature of the boundary, evaluated in a flat ambient space \cite{Witten:2024upt}.
\par The metric of the boundary and its normal vector are,
\begin{equation}
ds^2_{\partial\mathcal M}
= \left(1-\frac{r_s}{r}\right) d t_E^2 
  + r^2 d\Omega_{2}^2,
\end{equation}
and
\begin{equation}
n^\mu = \left(0,\sqrt{1-\frac{r_s}{r}},0,0\right),
\end{equation}
respectively. The Gaussian curvature associated to this submanifold is
\begin{equation}
    K(r) = \nabla_\mu n^\mu  = \frac{1}{\sqrt{g}}\partial_{\mu}(\sqrt{g}n^{\mu}) \nonumber
\end{equation}
\begin{equation}
= \frac{2\sqrt{1-\frac{r_s}{r}}}{r}
+ \frac{r_s}{2 r^2 \sqrt{1-\frac{r_s}{r}}}.
\end{equation}
\par Now we will compute $K_{0}$, the renormalization term, which is computed using a flat manifold with the same boundary,
\begin{equation}
    ds_{0}^{2} = \left(1-\frac{r_s}{r}\right) d t_E^2 + dr^{2},
  + r^2 d\Omega_{2}^2
\end{equation}
computing its Gaussian curvature,
\begin{equation}
    K_{0}(r) = \frac{2}{r}.
\end{equation}
\par Subtracting both Gaussian curvatures at $r=R_{M}$ and expanding in powers of $\frac{1}{R_{M}}$,
\begin{equation}
    K(R_{M})-K_{0}(R_{M}) = \frac{r_{s}}{2R^{2}_{M}}+ \mathcal{O}_{1}\Big(\frac{1}{R^{3}_{M}}\Big).
\end{equation}
Using this result and computing the Gibbons-Hawking-York action,

    \begin{equation}
    S^{R}[g_{\mu\nu}] = 4\pi GM^{2} +  \mathcal{O}_{2}\Big(\frac{1}{R_{M}}\Big)+F\Big(\frac{1}{R_{M}}\Big)
\end{equation}
\par Now by identifying $F\Big(\frac{1}{R_{M}}\Big)=-\mathcal{O}_{2}\Big(R_{M}^{-1}\Big)$, we have the contribution of the Euclidean Schwarzschild black hole to the partition function,
\begin{equation}
    S^{R}[g_{\mu\nu}] = 4\pi GM^{2}
\end{equation}.
\par An important remark is that it is not necessary to keep $r_{M}$ finite to find the correct contribution. We can see this by noticing that
\begin{equation}
\lim_{R_{M} \to \infty}
F\Bigg(\frac{1}{R_{M}}\Bigg)
= -\mathcal{O}_{2}\!\left(R_{M}^{-1}\right)
= 0 .
\end{equation}
We only kept $R_{M}$ finite in our calculations to explicitly show that the correct Gibbons-Hawking-York contribution can be evaluated using only compact manifolds.
\par The gravitational contribution to the vacuum amplitude, $4\pi G M^{2}$, arises from the classical Schwarzschild solution. Since this geometry is an on-shell vacuum solution of Einstein’s equations, the Einstein--Hilbert bulk action does not contribute, and the entire contribution comes from the Gibbons--Hawking--York boundary term, which encodes the boundary properties of the black hole spacetime. This result can be understood as an application of the stationary phase approximation \cite{gustafson2020mathematical}. Such an approach is widely used to describe tunneling processes in quantum mechanics and their generalization to topological tunneling phenomena in gauge theories \cite{Coleman_1985, Shifman:1994ee}.

\par Now that we have computed the semiclassical contribution, we will solve the path integral of the electromagnetic field in a Schwarzschild background. Due to the gauge invariance, this integral is singular \cite{WeinbergII, Fujikawa, Coleman_1985}. This issue is solved by fixing the gauge and introducing the Faddeev-Popov ghost fields,
\begin{equation}
    \mathcal{L'}(A_{\mu};c;\bar{c}) = \Big( -\frac{1}{4}F_{\mu\nu} F^{\mu\nu} +  \frac{1}{2\alpha}(\nabla^\mu A_\mu)^2 + \bar{c} \nabla^2 c\Big).
\end{equation}
Here we will work in the Feynman gauge, $\alpha=1$, where the components of $A_{\mu}$ decouple \cite{Fujikawa, Coleman_1985}.
\par Including the semiclassical contribution and the necessary gauge fixing, the partition function is 
\begin{equation}
     e^{-4\pi GM^2}\int \mathcal{D}\big[A_{\mu}, c,\bar{c}\big] \ e^{- \int d^{4}x \sqrt{g} \big(\frac{1}{2}A_{\nu}g^{\mu\sigma}\nabla_{\mu}\nabla_{\sigma}A^{\nu}+\bar{c}\nabla^{2}c \big)}.
\end{equation}
\par The path integral of free fields can be represented as functional determinants of the wave operators. Fields with bosonic and fermionic statistics lead to determinants with negative and positive powers, respectively.
\begin{equation}
    \mathcal{Z}(\beta_{H}) =\mathcal{N}e^{-4\pi GM^2}\big(\det (g_{\mu\nu}\nabla^{2}) \big)^{-\frac{1}{2}}\big(\det \nabla^{2} \big)^{1}=
\end{equation}
\begin{equation}
\mathcal{Z}(\beta_{H}) =\mathcal{N}e^{-4\pi GM^2}\big(\det \nabla^{2} \big)^{-1}  
\end{equation}
where the power $-1$ is the net result from contributions of the four components of the gauge and the two from the ghost fields, which describe the contributions from the two physical polarizations.
\par Functional determinants are current in quantum field theory in curved spacetimes. Here we will evaluate the functional determinant \cite{Fujikawa,Coleman_1985, Shifman:1994ee} using the operatorial zeta function 
 \cite{Hawking:1976ja, Coleman_1985,gustafson2020mathematical,Denef:2009kn}, of the Laplace-Beltrami operator, $\nabla^{2}=g^{\mu\nu}\nabla_{\mu}\nabla_{\nu}$, 
\begin{equation}
    \zeta_{\nabla^{2}}(s) = \Tr{\nabla^{2})^{-s}} = \sumint\frac{1}{(\lambda_{a})^{s}},
\end{equation}
where $\lambda_{a}$ are the eigenvalues of $\nabla^{2}$ and $\Tr$ is the trace. The trace can be a sum, in the case of the discrete spectrum, or an integral for the continuous ones; in other words, the topology of the spectrum determines if the trace is a sum, an integral, or both.
\par The relation between the zeta function, $\zeta_{\nabla^{2}}(s)$, and the functional determinant, $\det \nabla ^{2}$ is given by the beautiful formula,
\begin{equation}
  \zeta'_{\nabla^{2}}(0)  = -\ln \det \nabla^{2},
\end{equation}
which connects complex analysis, operator theory, and path integrals.

\section{Thermodynamics of Black Hole Atmospheres}
%%\label{}
\par Having the partition function $\mathcal{Z}(\beta)$ and using the Feynman-Kac theorem, we can compute the free energy of the system,
\begin{equation}
    \mathcal{F}(T_{H})= -\frac{1}{\beta_{H}}\ln \mathcal{Z}(T_{H})=
\end{equation}
\[ \text{GHY Action}
+\enskip\raisebox{0.5ex}{%
\feynmandiagram [inline=(a.base)] {
    --  a
  -- [photon, half left, edge label=$$] b 
  -- [photon, half left, edge label=$$] a,
};}
\enskip - \enskip 
\raisebox{0.2ex}{%
\feynmandiagram [inline=(a.base)] {
   -- a  
  -- [scalar, half left, edge label=$$] b 
  -- [scalar, half left, edge label=$$] a,,
};}
\]
\par Before delving into the calculation of the functional determinant, which gives the contribution from the vacuum fluctuations, we will first investigate the semiclassical gravitational contribution to the partition function of the electromagnetic vacuum in a Schwarzschild background, which comes from the boundary term,
\begin{equation}
    \mathcal{Z}_{BH} = \mathcal{N}e^{-\frac{\beta^{2}}{16G\pi^{2}}}=\mathcal{N}e^{-4\pi GM^{2}},
\end{equation}
which gives the free energy of the black hole,
\begin{equation}
    \mathcal{F}_{BH}(\beta) = \frac{M}{2},
\end{equation}
and also its famous black hole entropy, the Bekenstein entropy,
\begin{equation}
    S_{BH}(\beta) = 4\pi GM^{2}=\frac{A}{4G}. 
\end{equation}
We can check the consistency of our results by using the relation 
\begin{equation}
\mathcal{F}=\mathcal{U}-TS,
\end{equation}
which gives the expected result for the internal energy $\mathcal{U}=M$, its ADM mass. With these results, we can compute the specific heat of the black hole,
\begin{equation}
     C_{V}(T_{H})= -\frac{1}{8 \pi G T_H^2} < 0
\end{equation}
A negative heat capacity is a signature of instability. This happens because when a black hole emits radiation, its temperature increases.

\par Negative heat capacities are not exclusive to gravitational physics, but also present in molecular systems \cite{Schmidt:2000zs} and nuclei \cite{Moretto:2002gm, Borderie:2020oov}. For example, a cluster of sodium molecules can display a negative heat capacity \cite{Schmidt:2000zs}. Such phenomena occur due to small clusters of sodium having the tendency to convert added energy into potential, instead of kinetic energy. Thus, the additional energy is used to reconfigure the atoms instead of increasing their temperature. 
\par In quantum systems, negative heat capacities are not associated with general instabilities, but a signature of a system going through a first-order phase transition. For example, in the case of the Na clusters, there is a transition from solid to liquid \cite{Schmidt:2000zs}, and for the nuclei it is a liquid-gas \cite{Moretto:2002gm, Borderie:2020oov} transition; in both cases are first-order phase transitions. It is interesting to note that both systems, besides exhibiting a negative $C_{V}$, also exhibit the coexistence of different phases. In the end of this section we will discuss in more details the similarities between the physics of these very different systems. 
\par Here we will solve the determinant using the operatorial $\zeta$-function \cite{gustafson2020mathematical}. Thus, we need to solve the eigenvalue problem,
\begin{equation}
    \nabla^2 \Phi = \lambda \Phi,
\end{equation}
where $\Phi(\infty)=0$, and $\Phi(r_{s})$ is regular.
\par As the Schwarzschild has a large isometry group, $U_{T}(1) \times SO_{R}(3)$, time translations and rotations, respectively, we have three operators that commute with $\nabla^2 $,
\begin{equation}
    [\nabla^2;\hat{L}^{2}]=[\nabla^2;\hat{L}_{z}]=[\nabla^2;i\partial_{\mathbf{E}}]=0.
\end{equation}
Thus we can use the eigenvectors of these operators to build the eigenvectors of $\nabla^2$,
\begin{equation}
    \Phi(t_{\text{E}}, r, \theta, \phi) = e^{i\omega_{n} t_{\text{E}}} Y_{\ell}^{m}(\theta, \phi)\frac{\psi(r)}{r},
\end{equation}
where $\omega_{n}=\omega_{H}n=2\pi nT_{H}=\frac{n}{2r_{s}}$ are the Hawking thermal modes \cite{Soares:2019fed}, and the spherical harmonics appear due to the rotational invariance. Our eigenvalue problem can be further simplified by using tortoise coordinates \cite{carroll2003spacetime,hawking1973large},
\begin{equation}
    r_{*}(r) = r+r_{s}\ln\Big|{\frac{r}{r_{s}}-1}\Big|,
\end{equation}
resulting in a Schrödinger operator \cite{gustafson2020mathematical},
\begin{equation} \mathcal{L} \psi = \Big(-\frac{d^{2}}{dr_{*}^{2}} + V(r_{*})\Big) \psi = \lambda\psi \end{equation}
\begin{equation}
V_{\text{eff}}(r_{*}) = \frac{\ell(\ell+1)}{r_{}^2}\left(1 - \frac{r_{s}}{r}\right) + \frac{n^{2}}{4r^{2}_{s}} + \frac{(1-s^{2})r_{s}}{r^3}\left(1 - \frac{r_{s}}{r}\right),
\end{equation}
where $s=0,1$ and $2$ for scalar, vectorial and tensorial fields \cite{Hod:2016hdd}, respectively.
\par Summarizing, using the isometry group of the Schwarzschild spacetime and an adequate choice of system coordinates, we reduced the eigenvalue problem of the Laplace-Beltrami operator in $4D$ to an $1D$ Schrödinger operator.
\par Even for one--dimensional problems, very few exact solutions can be found. 
But fortunately, we do not need the detailed structure of the spectrum to study it. 
Given a Schr\"odinger operator, the analytical properties of $V_{\text{eff}}(r)$ are strongly related to its spectrum. 
Now we will use some properties of the effective potential $V_{\text{eff}}(r)$ that will be useful to characterize its spectrum.
\begin{enumerate}
  \item $V_{\text{Eff}}(r)\geq 0$, \\
  \item $V_{\text{Eff}}(r)$ is analytic in $(r_{s};\infty)$, \\
   \item $\lim_{r\to r_{s}}V_{\text{Eff}}(r)=\lim_{r\to +\infty}V_{\text{Eff}}(r)=\frac{n^2}{4r^{2}_{s}}$. 
\end{enumerate}
\par As $V(r)$ vanishes at infinity, the Weyl theorem \cite{gustafson2020mathematical} implies that the continuous spectrum, $\sigma_{c}(\mathcal{L})$ is non-empty and bounded from below by the thermal modes, in our case, $\sigma_{c}(\mathcal{L})=\left[\frac{n^2}{(2r_{s})^2}, +\infty \right)$. Given this information about the spectrum of $\nabla^{2}$, we can parameterize the eigenvalues as
\begin{equation}
\lambda_{n\mathbf{k}}=\frac{n^{2}}{4r_{s}^{2}}+\mathbf{\Gamma}^{2}, 
\end{equation}
where $n \in \mathbb{Z}$ and $\mathbf{\Gamma}^{2}$ is a real positive number related to the spatial contribution to the spectrum. As we are compressing the eigenvalues of a $3D$ problem in a $1D$ problem, each $\Gamma$ is highly degenerate. We can estimate this degeneracy by using the invariance under rotations of the system, we can write $\Gamma$ as an analytic function of $\mathbf{k}^2$, where $\mathbf{k} \in \mathbb{R}^3$. Considering only the leading order $\mathbf{\Gamma}^{2}(\mathbf{k}) \approx \mathbf{k}^2$, we have the following expression for the $\zeta_{\nabla^2}(s)$,
\begin{equation}
  \zeta_{\nabla^{2}}(s)= \frac{V}{(2\pi)^{3}}\sum_{n=-\infty}^{\infty}\int d^{3}k\frac{1}{\big[(\omega_{H} n)^{2}+\mathbf{k}^{2}\big]^{s}}.
\end{equation}
This expression can be simplified by noting that, in the sum over $n$, the contributions from $\pm n$ are identical, while the $n=0$ term is independent of the Hawking temperature $T_H$ and gives rise to an irrelevant divergence \cite{Hawking:1976ja, Percacci:2017fkn}. After discarding this temperature-independent contribution, we integrate by parts with respect to $k^{2}$ and perform a change of variables. This allows the expression to be factorized into the product of a sum and a integral.

\begin{equation}
   \zeta_{\nabla^{2}}(s)=- \frac{2V}{(2\pi^{2})(1+s)}\Big(2\pi T_{H}\Big)^{3-2s}\Big(\sum^{+\infty}_{n=1} n^{3-2s}\Big)\int_{0}^{+\infty} du\frac{1}{\big[1+u^{2}\big]^{s-1}}  \nonumber
\end{equation}
\begin{equation}
  = - \frac{V}{2\pi^{2}(1+s)}\Big(2\pi T_{H}\Big)^{3-2s} \zeta(2s-3)\frac{\Gamma\!\left(\tfrac{1}{2}\right)\Gamma\!\left(s-\tfrac{3}{2}\right)}{\Gamma(s-1)}. 
\end{equation}
where $\zeta(s)$ is the Riemann zeta function and $\Gamma(x)$ is the gamma function. Using the formula for the functional determinant and the Laurent series of $\zeta_{\nabla^{2}}(s)$ around $s=0$, we have the Helmholtz free energy of the photon gas,
\begin{equation}
    \mathcal{F}_{\text{A}}(T_{H},V)=
+\enskip\raisebox{0.5ex}{%
\feynmandiagram [inline=(a.base)] {
    --  a
  -- [photon, half left, edge label=$$] b 
  -- [photon, half left, edge label=$$] a,
};}
\enskip - \enskip 
\raisebox{0.2ex}{%
\feynmandiagram [inline=(a.base)] {
   -- a  
  -- [scalar, half left, edge label=$$] b 
  -- [scalar, half left, edge label=$$] a,,
};}
\end{equation}
\begin{equation}
=-\frac{4}{3}\sigma_{SB} T_{H}^{4}V_{\text{A}}
\end{equation}
where $\sigma_{SB}$ is the Stefan-Boltzmann constant and $V_{\text{A}}$ is the finite volume of the black hole quantum atmosphere. It is crucial to remark that the finiteness of the volume $V_{\text{Atm}}$ is not our imposition, but a necessary condition for the stability of the quantum atmosphere \cite{Gross:1982cv, Witten:2024upt}.  Roughly speaking, an arbitrarily small volume of gas at an arbitrary temperature may collapse into a black hole. Also, in a region very far from the event horizon, the vacuum will not be thermalized by the black hole; hence, the radiation must be concentrated in a region.
\par
The radius of the black hole quantum atmosphere can be estimated from the absorption cross section of high-energy photons scattered by a Schwarzschild black hole \cite{Fabbri:1975sa},
\begin{equation}
\sigma_{\text{abs}}\!\left(\omega \gg \frac{1}{r_s}\right)
= \frac{27\pi r_s^2}{4}
= \pi r_A^2 ,
\end{equation}
where, in the last equality, we identify the atmosphere radius $r_A$ as the effective radius in the eikonal limit,
\begin{equation}
r_A = \frac{3\sqrt{3}}{2}\, r_s \simeq 2.6\, r_s .
\end{equation}
This result implies that high-energy photons behave as if they were absorbed by a black hole of radius $r_A$. From a quantum perspective, these photons are absorbed by the thermalized quantum atmosphere surrounding the black hole.
\par Once the effective radius is known, the emitted power can be estimated as
\begin{equation}
P(T_H) = 27\pi \sigma_{\rm SB}\, r_s^2 \, T_H^4 .
\end{equation}
\par Now, considering both contributions to the vacuum energy, we have the free energy of the system,
\begin{equation}
    \mathcal{F}(T_{H},V)= \frac{M}{2} -\frac{4}{3}\sigma_{SB} T_{H}^{4}V_{\text{A}},
\end{equation}
\par which describes a Schwarzschild black hole surrounded by a cloud of photons at $T_{H}$. 
\par Using this free energy, we can evaluate the total entropy,
\begin{equation}
    S(T_{H}) =  \frac{A}{4G}+\frac{16}{3}\sigma_{SB}T^{3}V_{\text{A}},
\end{equation}
and the specific heat,
\begin{equation}
 C_{V}(T_{H})= -\frac{1}{8 \pi G T_H^2} + 16 \sigma_{SB} V_{\text{A}} T_H^3.
\end{equation}
which exhibits an intriguing change of sign at $T_{C}$, an indication that the system becomes stable at this temperature.
\begin{equation}
    T_{C}=\left( \frac{15}{256 \pi^3 G  V_{\text{A}}} \right)^{1/5}
\end{equation} 
\begin{figure}[H]
	\centering 
	\includegraphics[width=0.4\textwidth, angle=0]{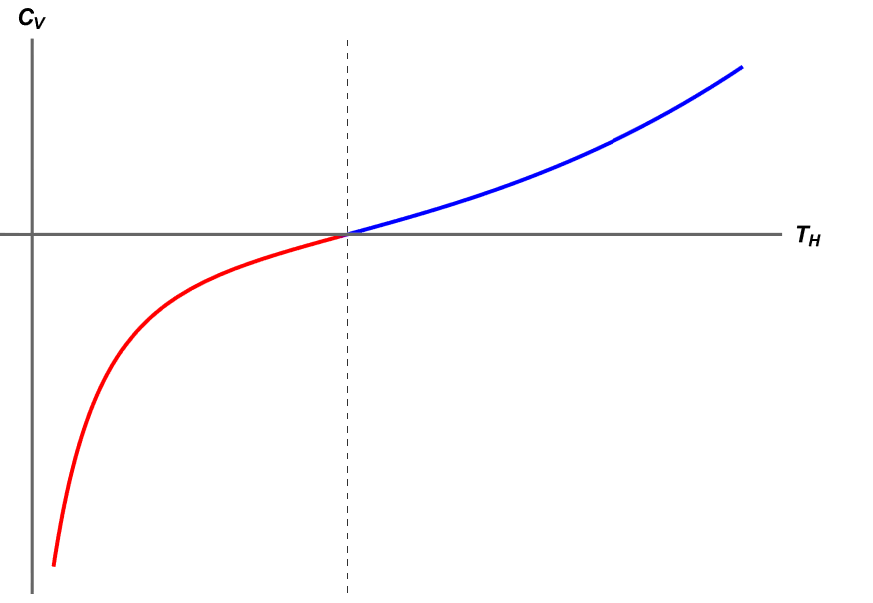}	
	\caption{This is the curve for the heat capacity of a black hole with the inclusion of the electromagnetic vacuum fluctuations. At $T_{C}$ the system reaches stability as $C_{V}>0$.} 
	\label{fig_mom03}%
\end{figure}
\par The reason for this change of sign is that as the black hole evaporates, more of its energy is transferred to the atmosphere, which gives a positive contribution to the specific heat. After a long time, the system will resemble an ordinary thermal quantum gas. This phase transition is similar to the Hawking-Page transition \cite{Hawking:1982dh, Giribet:2014fla, Braga:2025jji}, where a Schwarzchild-AdS black hole exhibits a competition between the black hole and cosmological event horizons. But in our case, $\Lambda=0$, so instead of a competition between two horizons we have a phase dominated by the black hole and another by the thermal gas. 
\par As explained earlier in this section, negative heat capacity is a signature of unstable systems undergoing first--order phase transitions with coexisting phases \cite{Schmidt:2000zs, Moretto:2002gm, Borderie:2020oov}. We may see the curve for $C_V(T)$ (illustrated in the Figure \ref{fig_mom03}) as a phase transition, where we have two phases coexisting: the black hole and its quantum atmosphere.

\par For quantum black holes, our formalism is not adequate because we are not taking into account the contribution of UV gravitons. In the case of astrophysical black holes, the contribution of gravitons can be estimated using the linearized Einstein equation, which yields a similar result, because photons and gravitons have the same number of polarizations. This can be seen as a clue that a black hole -- depending on its initial mass, charge, angular momentum -- may evaporate not completely, but until it becomes an ordinary thermalized body, without an event horizon.
\section{Corrections to the Entanglement Entropy of Black Holes}
\par A straightforward calculation shows a remarkable fact about the entropy of black holes. Given a star and a black hole, both with mass $M$, the entropy of a black hole is enormous compared to the entropy of a star, $S_{BH}\approx 10^{20}S_{\text{Star}}$. This is an indication that the process of the collapse of a star into a black hole increases the number of equivalent microstates, as black holes have enormous entropies when compared to stars. Therefore, a quantum theory of gravity should explain this anomalous increase in entropy.
\par Given an entropy, the natural question is what microstates such entropy is counting. Until now, the microscopic origin of Bekenstein entropy remains an open problem. The only microscopic derivation of the Bekenstein entropy has been found in the context of higher-dimensional supersymmetric black holes \cite{Strominger:1996sh}.
\par In recent years, approaches using techniques from quantum information have provided a fertile soil for new advances in quantum gravity. Here we will mention two: the proof of the Bekenstein bound by Horacio Casini \cite{Casini:2008cr}, and the Ryu-Takayanagi formula \cite{Ryu:2006bv}, a generalization of the entropy area law for the Von Neumann entropy of quantum fields, which has been particularly successful in the context of holography \cite{Bhattacharya:2012mi, Witten:2024upt}. Both results brought more physical insight about the generalized gravitational entropy \cite{Wald:1995yp, Lewkowycz:2013nqa, Almheiri2021},
\begin{equation}
    S_{\text{Gen}} = \frac{A}{4G} + S_{\text{Out}},
\end{equation}
which is a Von Neumann entropy \cite{Kudler-Flam:2023qfl}, the quantum version of the Shannon entropy \cite{Witten:2024upt}.
\par As general relativity is not renormalizable using perturbative methods, the calculation of $S_{\text{Out}}$ including gravitons gives a divergent result \cite{Kudler-Flam:2023hkl}. Due to this issue, we computed the correction to the entropy using quantum field theory in curved spacetimes, which yields a finite $S_{\text{Out}}$ \cite{Almheiri2021, Kudler-Flam:2023hkl},
\begin{equation}
    S_{\text{out}} = \frac{16}{3}\sigma_{SB}T^{3}V_{A}.
\end{equation}
Which is the contribution of the black hole atmosphere to the generalized gravitational entropy $ S_{\text{Out}}=S_{\text{A}} $.
Our result contains information about the interactions between gravitons and the emitted photons in the semiclassical limit \cite{Almheiri:2019hni}. Thus, in this limit, the entanglement entropy between the black hole and the electromagnetic vacuum, 
\begin{figure}[H]
	\centering 
	\includegraphics[width=0.2\textwidth, angle=0]{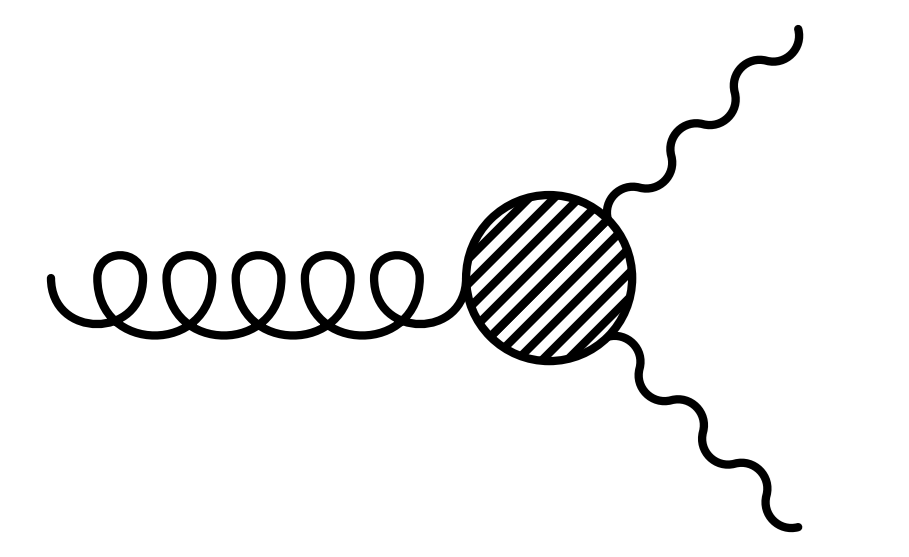}	
	\caption{In the particle-antiparticle picture of Hawking radiation, a pair is created due to the intense gravitational field; one of the particles goes into the black hole while the other is scattered to infinity. As electrodynamics does not possess self-interactions, the particle creation is only possible if gravitons are taken into account. In our approach, the gravitons are hidden in the background metric, but their effects appear in the observables. In this diagram, the blob represents loop corrections to the vertex. } 
	\label{fig_mom0}%
\end{figure}
\par The picture of Hawking radiation as the byproduct of the creation of a pair of particles-antiparticles near the event horizon is useful for gaining a better physical insight into our calculation of $S_{\text{Out}}$. Because this particle-antiparticle picture only holds for photons if gravitons are included (since electrodynamics is an Abelian gauge theory), the electromagnetic Hawking radiation must be the outcome of the decay of gravitons into photons, as illustrated in the Figure $\ref{fig_mom0}$. The semiclassical quantum correction to the entropy of a black hole gives an approximation of the entanglement between the electromagnetic vacuum and a Schwarzschild black hole. 

\section{RVB Formula and the Euler Characteristic of Black Holes}\
\par The idea of the RVB formula \cite{Robson:2018con} is to constrain the Hawking temperature to the Euler characteristic of the spacetime using the Gauss-Bonnet theorem for $2D$ manifolds,
\begin{equation}
    \chi(\mathcal{M})=\frac{1}{2\pi}\int_{\mathcal{M}}\mathcal{K}_{g} + \frac{1}{2\pi}\int_{\partial\mathcal{M}}k_{g},
\end{equation}
where $\mathcal{K}_{g}$ is the trace of the gaussian curvature of $\mathcal{M}$ and $k_{g}$ is the geodesic curvature of the boundary $\mathcal{\partial\mathcal{M}}$ \cite{Nakahara:2003}. This theorem, which is the main bridge between geometry and topology  $2D$ manifolds, is generalized for higher-dimensional manifolds by the Chern-Gauss-Bonnet theorem \cite{Nakahara:2003}. Although the latter can be applied to $D$-dimensional black holes, its calculation leads to more complicated integrals. To avoid this problem, here we will use algebraic topology \cite{Hatcher, Schwarz:1994ee}. Using these methods, the computation of these topological invariants is reduced to simple calculations in any dimension.
\par The Euler characteristic of a manifold can be evaluated using its homology groups, $H^{c}_{k}(\mathcal{M})$. For compact manifolds,
\begin{equation}
    \chi(\mathcal{M}) = \sum_{k=0}^{D}(-1)^{k}b_{k}(\mathcal{M}),
\end{equation}
where $b_{k}=\dim H^{c}_{k}(\mathcal{M})$, the Betti numbers of the manifold. These are natural numbers that encode information about the holes and other connectivity properties of the manifold. For $2-$manifolds these numbers have a clear interpretation: $b_{0}$ counts the number of connected pieces, $b_{1}$ the number of holes, and $b_{2}$ counts the two-dimensional holes (the number of ´´voids´´ or ''cavities'' in a manifold). 
\par Here we are using $H_k^c(M)$ -- the compactly supported homology \cite{Hatcher}- instead of $H_k(M)$, because we have to deal with $\mathbb{R}^2$, which is not a compact manifold.
\par A fundamental theorem of algebraic topology is the Künneth formula \cite{Hatcher,Nakahara:2003}, which states that the homology of the product is the product of the homologies. Applying it to our case,
\begin{equation}
 H^{c}_{k}( \mathbb{R}^{2} \times \mathbb{S}^{2})=   H^{c}_{k}(\mathbb{R}^{2}) \otimes  H_{k}( \mathbb{S}^{2}).
\end{equation}
\par This formula is useful to decompose a four-dimensional spacetime in smaller dimensional manifolds, in our case $\mathbb{D}^{2}$ and $\mathbb{S}^{2}$. This simplifies the calculation of the homology of black holes because the topology of two-dimensional manifolds is fully understood. In contrast, $4$-manifolds are an active area of research. This analysis holds for Kerr and RN BHs, which have the same topology, and it can be extended to other BHs.
\par The non-zero homology groups of $\mathbb{R}^{2}$ and $\mathbb{S}^{2}$, are $H^{c}_{0}(\mathbb{S}^{2})=H^{c}_{2}(\mathbb{S}^{2})=H^{c}_{0}(\mathbb{R}^{2})=\mathbb{Z}$ \cite{Hatcher,Nakahara:2003}. Now we can compute the Betti numbers using  $b_{k}=\dim H_{k}(\mathcal{M})$, given the following nonzero Betti numbers, 
\begin{equation}
     b_{0}(\mathbb{S}^{2})=b_{0}(\mathbb{D}^{2})=b_{2}(\mathbb{S}^{2})=\dim \mathbb{Z}=1.
\end{equation}     
Thus $\chi(\mathbb{S}^{2})=2$ and $\chi(\mathbb{R}^{2})=1$.  Using a corollary of the Kunneth formula, we have the Euler characteristics of the Schwarzschild spacetime, 
\begin{equation}
\chi({\mathbb{R}^{2} \times \mathbb{S}^{2})=\chi(\mathbb{R}^{2}})\chi(\mathbb{S}^{2}) = 1 \times 2= 2
\end{equation}
\par Now that we reduce the $4D$ problem to $2D$, the analytical methods are straightforward to apply. Using the Gauss-Bonnet theorem, we will show that there is a constraint between the observables of a black hole and the spacetime topological invariants.
\par The Gaussian curvature of the $2$-sphere, $\mathbb{S}^{2}$ is $\mathcal{K}_{g} = r^{-2}$, a direct application of the Gauss-Bonnet theorem gives
\begin{equation}
    \chi(\mathbb{S}^{2})=\frac{1}{2\pi}\int_{\mathbb{S}^{2}} d\theta d\varphi \sin{\theta} \ r^{2}\frac{1}{r^{2}}=2
\end{equation}
as expected. The evaluation of the other factor that will give the RVB formula.
\par The submanifold $\mathbb{R}^{2}$ is described by the metric,
\begin{equation}
    ds^{2} = \Big(1-\frac{r_{s}}{r}\Big)dt_{\text{E}}^{2}+\Big(1-\frac{r_{s}}{r}\Big)^{-1}dr^{2},
\end{equation}
a deformation of the usual disk with a Gaussian curvature
\begin{equation}
\mathcal{K}_{2}=\frac{r_{s}}{r^{3}}.
\end{equation}
\par Using the Betti numbers, $\chi(\mathbb{R}^{2})=1$, and the Gauss-Bonnet theorem, we have the Hawking temperature in terms of $\chi$.
\begin{equation}
  \chi(\mathbb{R}^{2})=\frac{1}{2\pi}\int_{\mathbb{R}^{2}}\ \mathcal{K}=\frac{1}{2\pi} \int_{0}^{\beta_{H}}dt_{E}\int^{\infty}_{r_{s}} dr  \frac{r_{s}}{r^{3}}= \nonumber
\end{equation}  
  \begin{equation}
 \chi\big(\mathbb{R}^{2}\big)=1=  \frac{\beta_{H}}{4\pi r_{s}},
\end{equation}
which is the original RVB formula for the Schwarzschild black hole \cite{Robson:2018con}.
\par Although it seems counterintuitive, it is not the contribution from the event horizon $\mathbb{S}^{2}$, the origin of the RVB formula, but the $\chi(\mathbb{R}^{2})$ that can be used to constrain the Hawking temperature. Sugesting that, in general, the quotient of the spacetime and its event horizon, 
\begin{equation}
    \mathcal{M}_{Q}=\frac{\mathcal{M}_{BH}}{\mathcal{M}_{EH}},
\end{equation}
that should be used in the RVB formula for $D$ dimensional spacetimes. 
\par To prove that this conjecture holds, we will compute the Hawking temperature of a $D$ dimensional  Schwarzschild-Taglerini black hole \cite{Emparan:2008eg}, which can be described by the following element line,
\begin{equation}
    ds^{2} = \Big(1-\frac{\mu^{D-3}}{r^{D-3}}\Big)dt^{2}_{\text{E}}+\Big(1-\frac{\mu^{D-3}}{r^{D-3}}\Big)^{-1}dr^{2}+r^{2}d\Omega^{2}_{\mathbb{S}^{D-2}},
\end{equation}
where $\mu$ is the horizon radius. Similarly to the usual Schwarzschild black hole, its topology is $\mathbb{R}^{2}\times \mathbb{S}^{D-2}$, where $\mathbb{S}^{D-2}$ is the event horizon. To compute its $\chi$, we use that $\chi(\mathbb{R}^{2})=1$, and the fact that spheres can have only two possible values for $\chi$: $\chi(\mathbb{S}^{D-2})=0$, if $D$ is odd, and $\chi(\mathbb{S}^{D-2})=1$ if $D$ is even. These calculations are detailed in \ref{AT}. Thus, the $\chi$ of the Schwarzschild-Taglerini black holes are 0 (2) if $D$ is odd (even). Along the same lines, the same is true for the Myers-Perry solutions \cite{Emparan:2008eg}.
\par Now applying the RVB formula to the Schwarzschild-Taglerini black hole, which will be given by the evaluation of
\begin{equation}
    \chi\Big(\frac{\mathbb{R}^{2}\times \mathbb{S}^{D-2}}{\mathbb{S}^{D-2}}\Big)= \chi(\mathbb{R}^{2})  =1.
\end{equation}
In the $D$-dimensional case, the deformation of the $\mathbb{R}^{2}$ associated to the quotient above is
\begin{equation}
    ds^{2} = \Big(1-\frac{\mu^{D-3}}{r^{D-3}}\Big)dt^{2}_{\text{E}}+\Big(1-\frac{\mu^{D-3}}{r^{D-3}}\Big)^{-1}dr^{2},
\end{equation}
by computing its Gaussian curvature we found,
\begin{equation}
    \mathcal{K}_{D} = \frac{(D-3)(D-2)\mu}{2r^{D-1}}.
\end{equation}
As in the $2D$ case, by applying the Gauss-Bonnet theorem, we find
\begin{equation}
    \frac{\beta}{4\pi}\frac{D-3}{\mu}=\chi(\mathbb{R}^{2})=1,
\end{equation}
hence,
\begin{equation}
    T_{H} = \frac{D-3}{4\pi \mu},
\end{equation}
the correct Hawking temperature of the $D-$dimensional Schwarzschild-Tangherlini black hole \cite{Emparan:2008eg, Hod:2016hdd}.
\section{Topological Tunneling Interpretation \\ of Black Hole Evaporation: Euclidean Black Holes as Instantons}
\par The relation between black hole thermodynamics and spacetime topology first was noticed in the investigation of gravitational instantons 
\cite{EguchiFreund1976,Hawking1978,Eguchi:1978xp,Gibbons:1979xm}. In Yang-Mills theories, instantons \cite{Belavin:1975fg,Shifman:1994ee} are classical solutions associated with quantum tunnellings between topologically inequivalent vacua. These solutions were particularly successful in QCD, solving the $U_{A}(1)$ Problem and in the discovery of the $\theta$ vacuum \cite{Jackiw76, Callan:1976je}, a vacuum characterized by a coherent state of topologically inequivalent (but degenerated in energy) vacua. Instantons -- and monopoles --  are classified using an integer number, the homotopy class of the gauge group \cite{Coleman_1985,Shifman:2012zz}, which is the topological charge of the instanton or monopole. Similarly, the vacua in Yang-Mills theory is also classified by these integer numbers. In the quantization, the vacuum amplitude is dominated by the contribution of a dilute gas of instantons, inducing the tunneling. In our case, the amplitude is dominated by the Gibbons-Hawking-York boundary term, and the black holes can be classified using their homology, as we show in the previous section. 
\par A black hole is regarded as a highly excited quantum gravitational state \cite{Bekenstein:1974jk,tHooft:1984kcu, Hod:1998vk,Corda:2019vuk}. We can see this by looking in the implications of the Positive Energy theorem \cite{Witten:1981mf, Gibbons:1982jg} -- a proof that the ADM energy cannot be negative -- and the No Hair theorem \cite{Israel:1967wq,hawking1973large}, which states that a black hole is completely characterized by three observables: mass, angular momentum, and charge. The Positive Energy Theorem also states that the Minkowski spacetime is the only solution with zero ADM energy. Also, we show that the Euler characteristics $\chi$ play a crucial role in black hole thermodynamics, as the RVB formula show. Thus, we can characterize the bulk of the gravitational ground state as 
\begin{equation}
\ket{\Omega_{Flat}}=
\ket{M=0,Q=0,J=0,\chi=1}.
\end{equation}
\par The role of $\chi$ as a quantum number for gravitational states also is useful is cosmological scenarios where $\Lambda\neq 0$. For example, in the case of a positive cosmological constant, $\Lambda >0$, the gravitational ground state is the de Sitter spacetime. As the Minkowski spacetime, de Sitter is a maximally symmetric spacetime, but with a more interesting topology, $\mathcal{M}_{dS}= \mathbb{R} \times \mathbb{S}^3$, hence $\chi=0$ and $\ket{\Omega_{\text{dS}}}=\ket{0;0;0;0}$. This result is particularly interesting because it shows that $\chi$ is also a suitable number to identify cosmological horizons and different gravitational ground states given an arbitrary value of $\Lambda.$
\par The Positive Energy Theorem also applies to black hole solutions. In particular, they give important inequalities between $M$, $Q$ and $J$ \cite{Gibbons:1982jg,Dain:2011mv}, 
\begin{equation}
    M^2 \geq \frac{Q^{2}+\sqrt{Q^{4}+4J^{2}}}{2}. \label{ineq}
\end{equation}
\par We can interpret this inequality using the analogy between hydrogen atoms and black holes \cite{Bekenstein:1974jk, Mukhanov:1986me, Hod:1998vk}. The quantum numbers of the hydrogen states, $\ket{n,l,m}$, are not independent; they must obey the inequalities: $n> \ell$ and $|m|\leq \ell $. Analogously, we can see the inequality $\ref{ineq}$ as a similar constraint between the black hole observables. 
\par
\begin{figure}
	\centering 
	\includegraphics[width=0.4\textwidth, angle=0]{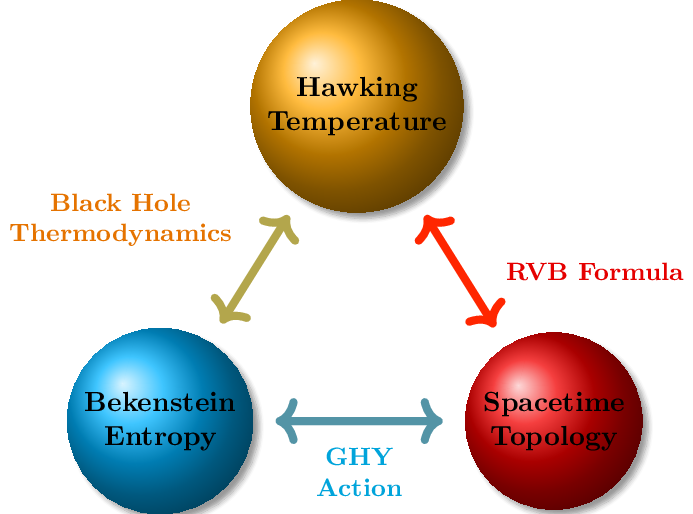}	
	\caption{This diagram illustrate the deep connections between spacetime topology and black hole thermodynamics. The relation between the Gibbons-Hawking-York action was found by S. Liberatti and others \cite{Liberati:1995jj, Liberati:1996kt, Liberati:1997sp} , and the connection between $\chi$ and $T_{H}$ was discovered by RVB \cite{Robson:2018con}. } 
	\label{fig_mom02}%
\end{figure}

 Besides $M$, $Q$, and $J$, the Euler characteristic, $\chi$, is also a good quantum number given its direct connection with the Hawking temperature \cite{Robson:2018con} and also with the black hole entropy through the Gibbons-Hawking-York action \cite{Liberati:1996kt, Liberati:1997sp}, as illustrated in Figure \ref{fig_mom02}. Hence, the bulk of a black hole state can be described by
\begin{equation}
\ket{\Omega_{BH}}=\ket{M,J,Q, \chi}
\end{equation}
where in the Schwarzschild case, $\ket{\Omega_{BH}}=\ket{M,0,0,\chi=2}$. In addition, the $\chi$ is necessary to indicate the existence of the event horizon \cite{galloway1993topology}. Without the $\chi$, the state does not describe a black hole, but any system with mass $M$. The Euler characteristic can be viewed as a topological quantum number related to the existence of event horizons.
\par The analogy between hydrogen atoms and black holes may seem flawed given a main difference: while the $n$, $\ell$, and $m$ are integer numbers, $M$, $Q$, and $J$ are real numbers. This issue is solved by looking at Rydberg atoms \cite{Wu:2020axb}, which are hydrogenoid states with principal quantum numbers of the order $n\sim 100$. These highly excited atomic states exhibit a nearly continuous spectrum, as the energy gap between two consecutive states is around $10^{-5}$ eV. Rydberg atoms are unstable and rapidly decay to their ground states by emitting photons. Along these lines, we can view a black hole as a sort of gravitational Rydberg atom, where the decay mechanism to its ground state is the emission of thermal radiation, known as the Hawking effect.  

\par  From the topological point-of-view,the black hole evaporation is a transition from the Schwarzschild to the flat spacetime.
\begin{equation}
\mathbb{S}^{2}\times \mathbb{R}^{2} \rightarrow \mathbb{R}^{4}. 
\end{equation}
Thus, the black hole evaporation is related to a change in the spacetime topological invariants, in particular, a variation in the Euler characteristics, as $\chi(\mathbb{S}^{2}\times \mathbb{R}^{2})=2$ and  $\chi(\mathbb{R}^{4})=1$. In this picture, evaporation is a direct result of the variation $\Delta \chi = -1$.
\begin{equation}
  \ket{M,0,0,\chi=2} \xrightarrow{\text{BH Evaporation}} \ket{M,0,0,\chi=1},
\end{equation}
where $\ket{M,0,0,1}$ represents an approximately flat spacetime -- and without an event horizon -- containing low density thermalized matter.
\par As the topology of Kerr-Newman, Schwarzschild, and any other asymptotically flat stationary black holes is the same, $\mathbb{R}^{2} \times \mathbb{S}^{2}$,  the evaporation of an asymptotically flat stationary 4D black hole is associated with $\Delta \chi =-1$. This result holds for $D$ even, for $D$ odd, $\Delta \chi =+1$. These calculations are explained in \ref{AT}.
\par These results lead to a new physical interpretation of Euclidean black holes as instantons; this class of solution represents a tunneling between topologically inequivalent spacetimes, in particular, the quantization of gravitational instantons describes the tunneling between spacetimes with different Euler characteristics. The variation of $\chi$ can be associated with the creation and destruction of event horizons.
\par There is an intriguing parallel between instantons in Yang-Mills and general relativity: instantons in Yang-Mills theories are related to variations in the homotopy class of the vacuum; in contrast, gravitational instantons alter the spacetime homological structure. This may indicate that the spacetime homology is relevant to describing general quantum gravitational states. Perhaps a more detailed quantum description of spacetime is given not only by $\chi$, but by all Betti numbers $b_{k}(\mathcal{M})$, which completely characterizes the homology of a manifold \cite{Hatcher, Nakahara:2003}.
\par An interesting question regarding whether $\Delta \chi$  is monotonic or not was raised by Professor Takanayagi \cite{privtaka}. As we show, the sign of $\Delta \chi $ varies with $D$ and can display both signs. But, given a fixed $D$, the variation of $\chi$ is not necessarily monotonic. For example, in the process
\begin{equation}
\mathbb{R}^{4} \rightarrow \mathbb{S}^{2}\times \mathbb{R}^{2}  \rightarrow \mathbb{R} \times \mathbb{S}^{3},
\end{equation}
the first tunneling is related to $\Delta \chi=+1$, and the second to $\Delta \chi =-2$. This hypothetical double tunneling process can be seem as creation of a stationary black hole ($\mathbb{R}^{4} \rightarrow \mathbb{S}^{2}\times \mathbb{R}^{2} $) and the second -- the most intriguing -- as a transition to the de Sitter spacetime ($\mathbb{S}^{2}\times \mathbb{R}^{2}  \rightarrow \mathbb{R} \times \mathbb{S}^{3}$), which means that, in this scenario, the final result of the black hole evaporation is an expanding Universe with $\Lambda>0$. The first tunneling is simply the collapse of a star into a black hole. Currently, we do not have a mechanism to associate the induction of a cosmological constant to the evaporation of black holes. So, the second tunneling is a speculation, even though there is no physical law that forbids such phenomena.

\section{Relation with the Parikh-Wilczek Tunneling}
\par The interpretation of the Hawking radiation as the result of a pair creation near the event horizon dates back to Hawking's pioneering works, although this pair creation is not explicit in the original calculations \cite{Hawking:1975vcx}. The first derivation of the Hawking radiation that describes it as a quantum tunneling process was found by Parikh and Wilczek \cite{Parikh:1999mf}. In their seminal work, they applied the WKB method \cite{gustafson2020mathematical} to the Klein-Gordon equation to compute the decay rate of tunnelings across the event horizon. This approach also works for general fields and black holes \cite{Angheben:2005rm, Erbin:2017zwo}.
\par The wave equation of a massless and spinless particle in the Schwarzschild spacetime (in this section we are not using the Euclidean formalism),
\begin{equation}
    \Box \phi =0,
\end{equation}
where $\Box$ is the Klein-Gordon operator in the Schwarzschild spacetime.
\par In the WKB method, we look for a solution of the kind \cite{gustafson2020mathematical},
\begin{equation}
    \phi(x) = \exp \big(S(x) +iT(x)\big)
\end{equation}
where $S(x)$ is related to the decay rate, and it is a solution of the Hamilton-Jacobi equation.
\par As we are interested in processes across the event horizon, the Gullstrand–Painlevé coordinates are a good choice of system of coordinates,
\begin{equation}
    ds^{2}= -\Big(1-\frac{r_{s}}{r} \Big)dt^{2}+dr^{2}+2\sqrt{\frac{r_{s}}{r}}dtdr+r^{2}d\Omega^{2},
\end{equation}
as the region $r=r_{s}$ is not singular. 
\par In the WKB approximation, the $s-$waves are associated with radial geodesics. Because waves with high angular momentum are more likely to inspiral back into the black hole, the bulk of the Hawking radiation arises from the contribution of the $s-$waves. Thus, to compute the evaporation of the black hole using the WKB approximation, we must use the following null radial geodesic, 
\begin{equation}
    \dot{r} = \pm 1-\sqrt{\frac{r_{s}}{r}},
\end{equation}
which leads to \cite{Parikh:1999mf,Angheben:2005rm, Erbin:2017zwo}
\begin{equation}
    \Gamma(\omega) \approx e^{-8\pi G\omega\big(M-\frac{\omega}{2}\big)}.
\end{equation}
\par Considering the case of an astrophysical black hole emitting quanta of energy, $\omega \ll M$,
\begin{equation}
    \Gamma(\omega \ll M) \approx e^{-8\pi GM\omega },
\end{equation}
which is the Boltzmann distribution for $T_{H}= \frac{1}{8\pi GM}$. Using this, we can recover the Stefan-Boltzmann law that we found in the Section 3 using path integrals.
\par Now looking for the total decay rate -- the complete evaporation of the black hole --,
\begin{equation}
    \Gamma(\omega=M) \approx e^{-4\pi GM^{2}} \approx \mathcal{Z}_{BH} = e^{-S_{GHY}},
\end{equation}
This shows that the total decay rate of the Parikh-Wilczek tunneling is exactly the contribution of the Gibbons-Hawking-York boundary term that we found using the stationary phase method.
\begin{equation}
    \Gamma(\omega=M) \sim e^{-4\pi GM^2}.
\end{equation}
The equivalence between the Parikh–Wilczek tunneling method and our Euclidean path integral formulation reinforces the tunneling interpretation of Hawking radiation. As shown in Section 7, the complete evaporation of the black hole is associated with a change in spacetime topology, characterized in particular by a variation of the Euler characteristic. This suggests that gravitational instantons play a role analogous to that of Yang–Mills instantons \cite{Shifman:1994ee}.

\section{Conclusions and Final Remarks}
%%\label{}

\par Here for the first time, we predicted the existence of the quantum black hole atmospheres using the Euclidean path integral quantization of the electromagnetic field near a Schwarszchild black hole. The prediction of these atmospheres was made by using the canonical quantization applied to scalar fields \cite{Giddings:2015uzr, Dey:2017yez, Dey:2019ugf, Ong:2020hti}. Our result for the free energy describes the black hole surrounded by a finite volume of photons at $T_{H}=\frac{1}{8\pi G M}$, the black hole quantum atmosphere. We computed the contribution of this atmosphere to the entropy -- which may be interpreted as a semiclassical approximation to entanglement entropy \cite{Almheiri2021} -- and to the specific heat. We find a positive contribution to the specific heat, which dominates above a critical temperature $T_C$, leading to a transition to $C_V(T_H)>0$. Together with the well-known connection between negative heat capacity and first-order phase transitions \cite{Schmidt:2000zs, Moretto:2002gm, Borderie:2020oov}, this suggests that black hole evaporation may be interpreted as a phase transition.

\par To further understand the role of spacetime topology in black hole thermodynamics, we found a new proof of the RVB formula, which relates the Euler characteristics, $\chi$, to the Hawking temperature. The topological invariant $\chi$ can be seem as a quantum number related to the existence of event horizons. It is useful to distinguish the states of a star $\ket{M;Q;J;\chi=1}$ and black hole, $\ket{M;Q;J;\chi=2}$ with the same masses. The topological quantum number $\chi$ fits the picture of a black hole as a sort of highly excited "atom", as originally proposed by J. Bekenstein \cite{Bekenstein:1974jk}. We argued here that the black hole evaporation can be seem as topological tunneling between spacetimes with different $\chi$. This result endorses the interpretation of black holes (and other exact solutions) as gravitational instantons, given the similarity with the quantization of Yang-Mills in a instantonic background, where the instantons induces tunnelings between topologically inequivalent vaccua. Besides the variation of $\chi$ after the complete evaporation, the RVB formula \cite{Robson:2018con} and the connection between black hole entropy, Gibbons-Hawking-York action and $\chi$ found by S. Liberati and G. Pollifrone \cite{Liberati:1997sp} are other indications of the central role of the spacetime topology in the evaporation of black holes.
\par The method presented here can be straightforwardly generalized to massive and interacting fields, including Yang--Mills theories, in the Schwarzschild background. For Reissner--Nordstr\"om black holes, only minor modifications are required, as the isometry group remains unchanged. The Kerr case, which is of particular astrophysical interest, is more involved, since the $SO(3)$ rotational symmetry is reduced to axial symmetry, $U(1)$. This reduction of the isometry group prevents the dimensional reduction of the four-dimensional Laplacian to an effective one-dimensional Schr\"odinger operator.
\par The detection of gravitational waves from black hole mergers and the first direct images of black holes have renewed interest in the pursuit of a quantum theory of gravity. With continued experimental progress and increasingly precise data, it may become possible in the near future to directly test the prediction of Hawking radiation and its consequences. The methods developed here provide a systematic framework to compute quantum corrections to classical black hole spacetimes arising from quantum fields near the event horizon. Given the central role of black holes in quantum gravity, we expect these techniques to be valuable tools for probing deviations from classical general relativity induced by quantum effects.
\section*{Acknowledgements}\par The author would like to thank Carlos Zarro and Gabriel Picanço for valuable discussions and detailed comments on the first draft of this manuscript. The author also thanks the mathematicians Alejandro Cabrera and Antonio Mac Dowell for their help with homology groups. The author is also grateful to Gaston Giribet and Tadashi Takanayagi for insightful comments related to the poster presentation of this work at the Giambiagi School (2025). This study was financed in part by the Coordenação de Aperfeiçoamento de Pessoal de Nível Superior – Brasil (CAPES) — Finance Code 001.
\appendix

\section{Euclidean Schwarzschild Black Hole}
\label{Eu}
\par Outside the event horizon, a Schwarzschild black hole can be described using spherical coordinates,
\begin{equation}
    ds^{2} = -\Big(1-\frac{r_{s}}{r}\Big)dt^{2}+\Big(1-\frac{r_{s}}{r}\Big)^{-1}dr^{2} + r^{2}d\Omega^{2}_{2}
\end{equation}
or using the Kruskal-Szekeres coordinates \cite{carroll2003spacetime,hawking1973large},
\begin{equation}
    ds^{2} = \frac{4r^{3}_{s}}{r}e^{-\frac{r}{r_{s}}}dUdV + r^{2}d\Omega^{2}_{2}
\end{equation}
where the relation between these two charts -- system of coordinates -- is given by
\begin{equation}
 U= \sqrt{\dfrac{r}{2GM}-1}\,e^{\frac{r}{2r_{s}}}\cosh\left(\frac{t}{2r_{s}}\right), \
\end{equation}
and,
\begin{equation}
 V = \sqrt{\dfrac{r}{2GM}-1}\,e^{\frac{r}{2r_{s}}}\sinh\left(\frac{t}{2r_{s}}\right), 
\end{equation}
which leads to,
\begin{equation}
    \frac{V}{U} = \tanh\left(\dfrac{t}{2r_{s}}\right).  
\end{equation}
\par Now we will study how the transition map between the two charts behaves under Wick rotation. This calculation demonstrates that the compactification of Euclidean space is necessary to ensure the metric is indeed a smooth solution of the Euclidean Einstein equations.
\par One of the axioms of smooth manifolds is that given two charts, the transition map -- the Jacobian matrix -- between them must be analytic (holomorphic) if the manifold is real (complex)\cite{Nakahara:2003}. In our case, this only holds if we use the standard analytic continuation of the hyperbolic trigonometric functions,
\begin{equation}
    \sinh\left(\frac{t}{2r_{s}}\right) \to  \sinh\left(\frac{it_{\textbf{E}}}{2r_{s}}\right)= i\sin\left(\frac{t_{\textbf{E}}}{2r_{s}}\right) \nonumber
\end{equation}
\begin{equation}
    \cosh\left(\frac{t}{2r_{s}}\right) \to  \cosh\left(\frac{it_{\textbf{E}}}{2r_{s}}\right)= \cos\left(\frac{t_{\textbf{E}}}{2r_{s}}\right) 
\end{equation}
Using basic properties of trigonometric functions, we find the period,
\begin{equation}
    t_{\textbf{E}} \sim  t_{\textbf{E}} + 4\pi r_{s}.
\end{equation}
Classically, this means that the Euclidean Schwarzschild solution must be compact in this imaginary direction. From the quantum perspective, a compact imaginary time is a signature of a system at a temperature,
\begin{equation}
    T_{H} = \frac{1}{4\pi r_{s}},
\end{equation}
the Hawking temperature.

\section{Algebraic Topology and Black Holes}
\label{AT}
\par \ \ \ Topology is concerned with global properties of spaces. In general relativity, methods from differential topology -- the investigation of global properties of manifolds using tools from analysis --- were essential in the revolution of Black Hole physics during the 60's and 70's \cite{hawking1973large}. During the same period, another branch of topology became widely present in particle physics: algebraic topology \cite{Coleman_1985,Shifman:1994ee,Nakahara:2003,Schwarz:1994ee}. The idea of this field is to study global properties of manifolds using abstract algebra, in particular, groups.
\par Given a manifold, its holes and connecteness properties are given by its homology, cohomology and homotopy groups \cite{Hatcher,Schwarz:1994ee,Nakahara:2003}. Although the study of these groups can lead to remarkably difficult problems for certain manifolds, here we are interested in well-understood spaces, such as $\mathbb{S}^{2}$, $\mathbb{T}^{2}$($\mathbb{S}^{1}\times\mathbb{S}^{1}$), and $\mathbb{R}^{2}$. Here we are interested in the homology groups, and as the later is a non-compact manifold, we must use the compactly supported homology groups, $H^{c}_{n}(\mathcal{M})$ instead of the usual homology groups, $H_{n}(\mathcal{M})$. 
\par Below a table with the homology groups and $\chi$'s of some manifolds of interest,
\begin{table}[H]
\centering
\begin{tabular}{l c c c c} 
 \hline
 $\mathcal{M}$ & $H^{c}_{0}(\mathcal{M})$ & $H^{c}_{1}(\mathcal{M})$ & $H^{c}_{2}(\mathcal{M})$ & $\chi(\mathcal{M})$ \\ 
                \\
 \hline
 $\mathbb{R}^{2}$ & 	$\mathbb{Z}$ & 0 & 0 & 1 \\ 
  $\mathbb{S}^{2}$ & 	$\mathbb{Z}$ & 0 & $\mathbb{Z}$ & 2  \\ 
 $\mathbb{T}^{2}$ & 	$\mathbb{Z}$ & $\mathbb{Z}\times \mathbb{Z}$ & $\mathbb{Z}$ & 0  \\ 
  \hline
\end{tabular}
\label{Table1}
\caption{This table contains the homology groups and the Euler characteristics of the $2$-dimensional compact that are useful in the context of black hole physics. The $\chi$ was evaluated using the alternate sum of $\dim H_{k}$}
\end{table}
Besides these two-dimensional manifolds, the spaces $\mathbb{S}^n$ and $\mathbb{H}^n$ also commonly appear in gravitational settings. In general, the sphere $\mathbb{S}^n$ has only two nontrivial homology groups,
\begin{equation}
H_0(\mathbb{S}^n) = H_n(\mathbb{S}^n) = \mathbb{Z},
\end{equation}
implying that its Euler characteristic satisfies $\chi(\mathbb{S}^n)=0$ for odd $n$ and $\chi(\mathbb{S}^n)=2$ for even $n$. For the non-compact space $\mathbb{R}^n$, ordinary homology yields
\begin{equation}
H_0(\mathbb{R}^n) = \mathbb{Z},
\end{equation}
with all higher homology groups trivial, giving $\chi(\mathbb{R}^n)=1$. However, when compactly supported homology \cite{Hatcher}  is considered, $\mathbb{R}^n$ possesses a single nontrivial group,
\begin{equation}
H_n^{c}(\mathbb{R}^n) = \mathbb{Z},
\end{equation}
which leads to the compactly supported Euler characteristic
\begin{equation}
\chi_c(\mathbb{R}^n) = (-1)^n .
\end{equation}

\bibliographystyle{ieeetr}
\bibliography{example}



%% For example:

%% \bibitem[Aladro et al.(2015)]{Aladro15} Aladro, R., Martín, S., Riquelme, D., et al. 2015, \aas, 579, A101

%%\end{thebibliography}

\end{document}